\documentclass{aastex63}
\usepackage{color}
\usepackage{bm}
\usepackage{cases}
\usepackage{multirow}
\usepackage{lineno}

\received{}
\revised{}
\accepted{}
\submitjournal{}
\shorttitle{}
\shortauthors{}
\begin{document}

    \title{Gamma ray burst constraints on cosmological models from the improved Amati correlation}

    \author{Yang Liu}
      \affiliation{Department of Physics and Synergistic Innovation Center for Quantum Effects and Applications, Hunan Normal University, Changsha, Hunan 410081, China}

    \author{Nan Liang}
   \affiliation{Key Laboratory of Information and Computing Science Guizhou Province, Guizhou Normal University, Guiyang, Guizhou 550025, China}
   \affiliation{Joint Center for FAST Sciences Guizhou Normal  University Node, Guiyang, Guizhou 550025, China}
   
   \author{Xiaoyao Xie}
   \affiliation{Key Laboratory of Information and Computing Science Guizhou Province, Guizhou Normal University, Guiyang, Guizhou 550025, China}
   \affiliation{Joint Center for FAST Sciences Guizhou Normal  University Node, Guiyang, Guizhou 550025, China}
   
   \author{Zunli Yuan}
  \affiliation{Department of Physics and Synergistic Innovation Center for Quantum Effects and Applications, Hunan Normal University, Changsha, Hunan 410081, China}

    \author{Hongwei  Yu}
    \affiliation{Department of Physics and Synergistic Innovation Center for Quantum Effects and Applications, Hunan Normal University, Changsha, Hunan 410081, China}

    \author{Puxun Wu}
    \affiliation{Department of Physics and Synergistic Innovation Center for Quantum Effects and Applications, Hunan Normal University, Changsha, Hunan 410081, China}

    \email{yangl@hunnu.edu.cn}
    
   \email{liangn@bnu.edu.cn}
   \email{xyx@gznu.edu.cn}
   \email{hwyu@hunnu.edu.cn}
    \email{pxwu@hunnu.edu.cn}


\begin{abstract}
An improved Amati correlation was constructed in (ApJ 931 (2022) 50) by us recently. In this paper, we further study  constraints on the $\Lambda$CDM and $w$CDM models from the gamma ray bursts (GRBs) standardized with the standard and improved Amati correlations, respectively. 
By using the Pantheon type Ia supernova sample to calibrate the latest A220 GRB data set, the GRB Hubble diagram is obtained model-independently. 
We find that at the high redshift region ($z>1.4$) the GRB distance modulus from the improved Amati correlation is larger apparently than that from the standard Amati one. 
The GRB data from the standard Amati correlation only give a lower bound limit  on the present matter density parameter  $\Omega_{\mathrm{m0}}$, while the GRBs from the improved Amati correlation constrain the $\Omega_{\mathrm{m0}}$ with the $68\%$ confidence level  to be $0.308^{+0.066}_{-0.230}$ and $0.307^{+0.057}_{-0.290}$ in the $\Lambda$CDM and $w$CDM models, respectively, which are consistent very well with those given by other current popular observational data including BAO, CMB and so on. Once the $H(z)$ data are added in our analysis, the constraint on the Hubble constant $H_0$ can be achieved. We find that  two different correlations provide slightly different $H_0$ results but the marginalized mean values seem to be close to that from the Planck 2018 CMB observations.

\end{abstract}

\section{Introduction}
Current observations, such as the type Ia supernovae (SN Ia)~\citep{Riess1998,Perlmutter1999}, the cosmic microwave background radiation (CMB)~\citep{CMB1,CMB2}, and the baryon acoustic oscillation(BAO)~\citep{BAO}, indicate that the cosmic expansion is accelerating.
To explain the origin of this peculiar phenomenon, commonly, a perfect fluid with negative pressure in the Universe  dubbed dark energy  is introduced. The nature of dark energy can be characterized by its equation of state (EoS)   $w$. 
The simplest candidate of dark energy is the cosmological constant $\Lambda$, whose  EoS is $w=-1$. The cosmological constant dark energy plus the cold dark matter make up the $\Lambda$CDM cosmological model, and the $w$CDM model is obtained  if  the EoS of dark energy is generalized from $-1$ to an arbitrary  constant $w$. 
The simple $\Lambda$CDM model fits with observations \citep{eBOSS2021}  very well on one hand; however, on the other hand,  it is plagued by  the Hubble constant ($H_0$) tension \citep{Perivolaropoulos2021,Valentino2021}.
Based on the $\Lambda$CDM model, the high-redshift CMB data \citep{Planck} give a tight constraint on $H_0$ ($67.4\pm 0.5~\mathrm{ km~s^{-1} Mpc^{-1}}$),  which  deviates from $H_0=73.2\pm 1.3~\mathrm{ km~s^{-1} Mpc^{-1}}$ given by the low-redshift SN Ia data \citep{Riess2018a,Riess2018b,Riess2021} more than 4$\sigma$.
Although many other observations such as the Hubble parameter $H(z)$ measurements, the BAO, and the strong gravitational lenses have been used to discuss this $H_0$ tension, its origin   is not yet  determined \citep{Wu, Chen2017,Abbott2018,Birrer2020,Cao2021a, Lin2021,Khetan2021, Efstathiou2020,Freedman2021,Cao2022b}.
Since the farthest redshift of the popular SN Ia, $H(z)$ and BAO data is about $2$, while the CMB data lie near the redshift of 1100, the cosmological data during the region of  middle redshift ($2\lesssim z\lesssim1100$) may play an important role in understanding the origin of the $H_0$ tension. 

 Gamma ray bursts (GRBs) are extremely energetic and thus detectable at a redshift  up to $z\sim9.4$ \citep{Cucchiara2011}.
This implies that the GRBs have potential to serve as a new cosmological probe to the cosmic evolution in  the middle redshift region.
To standardize the GRBs and then use them to constrain the cosmological models, many  empirical correlations  between parameters of the light curves and/or spectra with the GRB luminosity or energy  have been proposed~ \citep{Norris2000, Fenimore2000,Amati2002,Yonetoku2004,Ghirlanda2004,Liang2005,Firmani2006,Dainotti2008,Dainotti2016,Demianski2017,Wang2017,Wang2022,Hu2021,Luongo2021,Muccino2021,Cao2022c,Dainotti2022}.  However, 
to  calibrate GRB correlations,  a cosmological model is usually applied and the resulting correlations are later used to probe the cosmic evolution. Thus the GRB cosmology suffers  the  so called  \textit{circularity} problem~\citep{Ghirlanda2006,Wang2015}.
To avoid this problem, two different methods have been proposed.
The first one is the \textit{simultaneous fitting} or \textit{global fitting} method~\citep{Ghirlanda2004b,Li2008}, in which   the coefficients of the correlations and the parameters of a cosmological model are constrained simultaneously from the GRB observations.
The second one is the \textit{low-redshift calibration}~\citep{Liang2008, Liang2010, Kodama2008, Wei2009}, which bases on the idea of distance ladder,  as the GRB correlations are calibrated  by using other low redshift observations   such as the $H(z)$ data \citep{Amati2019,Montiel2021} or the SN Ia data \citep{Demianski2017,Demianski2017b,Demianski2021}. 
Up to now, the GRB data have been used widely to explore the components of our Universe, the nature of dark energy, and the  Hubble constant tension \citep{Demianski2017,Demianski2017b,Demianski2021,Liu2015,Wang2016,Lin2016,Amati2019,Khadka2020,Cao2021b,Cao2022a}.

Among the GRB empirical correlations, the Amati correlation is a very   popular one, which connects the spectral peak energy in the GRB cosmological rest-frame and the isotropic equivalent radiated energy ($E_p-E_\mathrm{iso}$)~\citep{Amati2002,Amati2006a,Amati2006b,Amati2008,Amati2009,Amati2013}.  Recently, we proposed an improved Amati correlation \citep{Liu2022}  by using the Gaussian \textit{copula}
which is a powerful statistical tool capable of describing the dependence structures between multivariate random variables, and has  been applied  to various fields by the astronomical community \citep{Benabed2009,Koen2009,Jiang2009,Scherrer2010,Takeuchi2010,Yuan2018,Qin2020,Takeuchi2020}. In  \citep{Liu2022}, by choosing the spatially  flat $\Lambda$CDM model with $\Omega_{\mathrm{m0}}=0.30$ and $H_0=70~ \mathrm{ km~s^{-1} Mpc^{-1}}$ as the fiducial model, we utilize the low-redshift ($z<1.4$) GRB data to calibrate  the standard  and improved Amati correlations, and then extrapolate the results to the high-redshift GRB data to achieve the GRB Hubble diagram, where $\Omega_{\mathrm{m0}}$ is the present dimensionless matter density parameter. 
Using these calibrated GRBs to constrain the flat $\Lambda$CDM model, we found that the improved Amati correlation can give results well consistent with the fiducial model, while the standard one can not.
Thus, in \citep{Liu2022}, the reliability of the improved Amati correlation was ascertained with  a fiducial model, but its cosmological application  was not carried out. 
In this work, we will fill this gap.
In order to obtain   the  Hubble diagram of the latest A220 GRB samples \citep{Khadka2021} model-independently, we use  the Pantheon SN Ia data \citep{Scolnic2018} to calibrate the standard and improved Amati correlations, and then use these calibrated GRB data to constrain the $\Lambda$CDM and $w$CDM models. Besides the GRB data, the $H(z)$ data set is also added in our analysis to obtain a tight constraint on model parameters.  

The rest of the paper is organized as follows: Section \ref{Sec:Correlation&calibration} introduces the improved Amati correlation briefly, and standardizes the GRB samples by using the method of the low-redshift calibration. Section \ref{Sec:Cosmo&data} studies the constraints on  the $\Lambda$CDM and $w$CDM models from the GRB data and the $\mbox{GRB}+H(z)$ data. Our conclusions are summarized in Section \ref{Sec:Conclusion}.

\section{GRB Hubble diagram from low-redshift calibration}\label{Sec:Correlation&calibration}
\subsection{Improved and standard Amati correlations}

The standard Amati correlation is proposed by \cite{Amati2002}, which describes a  correlation between the spectral peak energy  $E_p$ and the isotropic equivalent radiated energy $E_\mathrm{iso}$ and has the form
\begin{equation}\label{Amati}
	y_\mathrm{Amati}=a+bx
\end{equation}
with
\begin{eqnarray}\label{epeisoxy}
	y\equiv \log\frac{E_\mathrm{iso}}{1\mathrm{erg}},~x\equiv \log\frac{E_p}{300\mathrm{keV}}.
\end{eqnarray}
Here ``$\log$'' denotes the logarithm to the base of 10, and $a$ and $b$ are free coefficients.
The spectral peak energy $E_p$ and the isotropic equivalent radiated energy $E_\mathrm{iso}$ in Eq.~(\ref{epeisoxy}) can be obtained through
\begin{eqnarray}\label{Ep}
	E_p&=&E_p^\mathrm{obs}(1+z),\\  \label{Eiso}
	E_\mathrm{iso}&=&4\pi d^2_L(z)S_\mathrm{bolo}(1+z)^{-1}, 
\end{eqnarray}
when the luminosity distance $d_L(z)$ is known. Here $E_P^\mathrm{obs}$ is the observed peak energy of the GRB spectrum,  and $S_\mathrm{bolo}$ is the bolometric fluence, which is an observable.

The improved Amati correlation is derived from the \textit{copula} function \citep{Liu2022}. 
To construct the correlation between $x$ and $y$ from the Gaussian \textit{copula}, $x$ and $y$ are  assumed to obey the Gaussian distributions, which are represented as $f(x)$ and $g(y)$, respectively. In addition,  the GRB is assumed to obey a special  redshift distribution  $w(z)=ze^{-z}$~\citep{Wang2017}. Then, we use $F(x)$,  $G(y)$  and $W(z)$ to indicate the cumulative distribution functions of  $f(x)$, $g(y)$ and $w(z)$, respectively.  According to the Sklar's theorem \citep{Nelson2006}, 
 a joint distribution function $\bar{H}$ can be constructed  by using the Gaussian \textit{copula} $C$:
\begin{eqnarray}
	\bar{H}(x,y,z;\bm{\theta})=C\left (F(x),G(y),W(z);\bm{\theta} \right).
\end{eqnarray}
Here $\bm{\theta}$ denotes the parameters of the \textit{copula} function. From the   density function of the joint distribution, we obtain  the improved Amati correlation, which takes the form~\citep{Liu2022}
\begin{eqnarray}\label{3_Amati}
	y_{\mathrm{copula}}&=& a+b\, x+c\, \mathrm{erfc}^{-1}[2 W(z)].
\end{eqnarray}
Here $\mathrm{erfc}^{-1}$ is the inverse of complementary error function and 	$W(z)=1-e^{-z} (1+z)$.
Clearly, the improved Amati correlation has an extra redshift-dependent term compared with the standard one given in Eq.~(\ref{Amati}). When the coefficient $c$ is fixed to be zero, the improved Amati correlation reduces to the standard one.

\subsection{  low-redshift calibration and GRB Hubble diagram}

Eq.~(\ref{Eiso}) indicates that  to  obtain the  coefficients in the correlations  from observations the luminosity distance needs to be given. In \citep{Liu2022}, this distance is given from the $\Lambda$CDM model and thus the results are model-dependent. To achieve a model-independent GRB Hubble diagram, here we utilize the method of the low-redshift calibration, and derive the luminosity distance in Eq.~(\ref{Eiso})  from the Pantheon SN Ia samples  by using the idea that at the same redshift the GRB has the same luminosity distance as the SN Ia.

\begin{figure}
	\centering
	\includegraphics[width=0.6\textwidth]{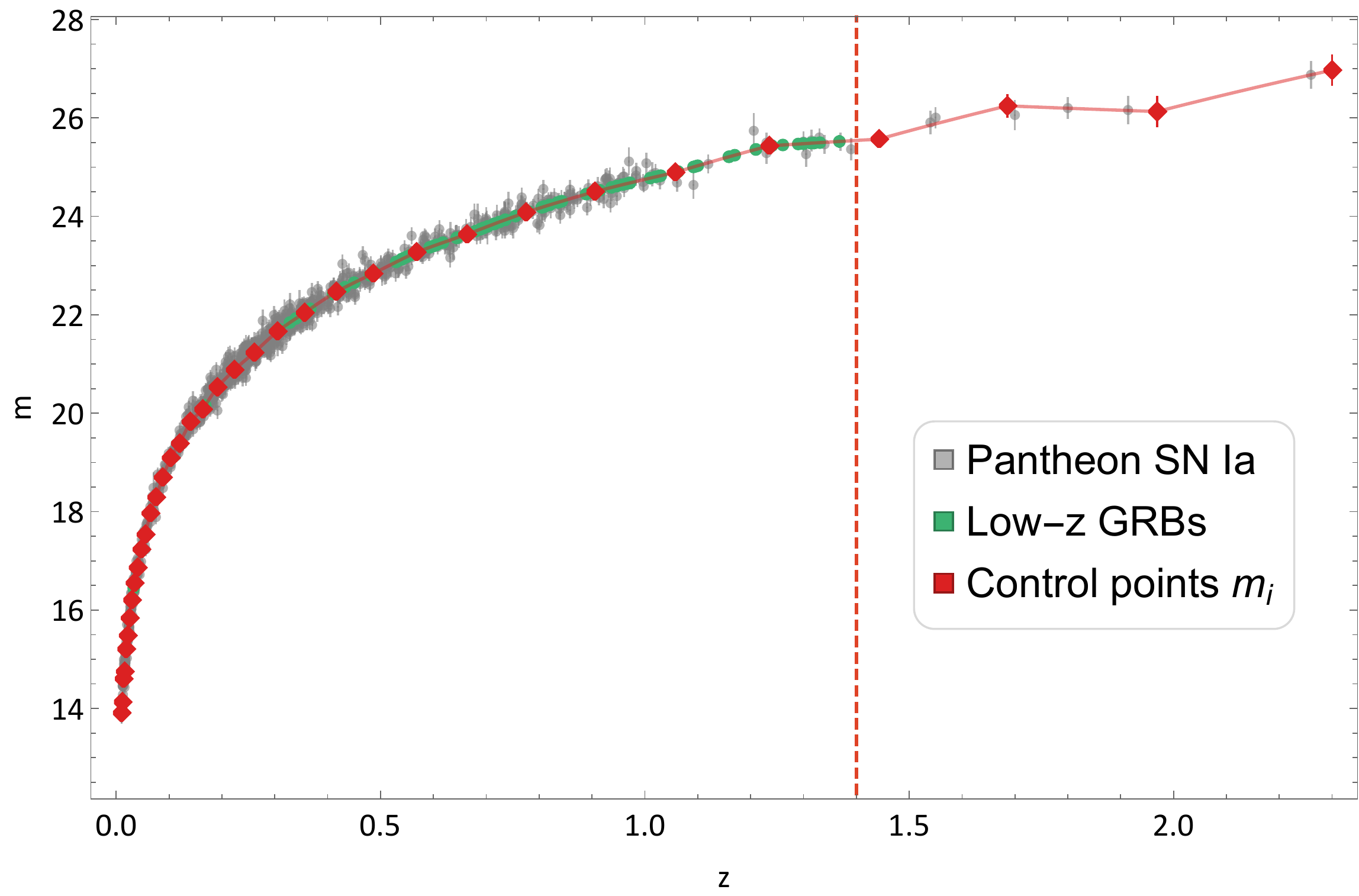}
	\caption{
		Apparent magnitudes of Pantheon SN Ia (gray points) and 36 $\log$-spaced control points (red points). The green points are low-redshift ($z<1.4$) GRBs calibrated by SN Ia. The red dashed line denotes $z = 1.4$.
		\label{fig1}
	}
\end{figure}

In our analysis, we consider the A220 data set,  which consists of A118 and A102 data sets and contains 220 long GRBs spanning  the redshift from 0.03 to 8.2~\citep{Khadka2021,Wang2016,Fana Dirirsa2019,Demianski2017,Amati2019}. This GRB data set is divided  into the low-redshift part ($z<1.4$), which has 79 data points, and the high-redshift one. 
To determine the values of coefficients in Eqs.~(\ref{Amati}) and (\ref{3_Amati}), we use the linear interpolation to estimate the luminosity  distance  of 79  low-redshift GRB data points  by following the method given in Appendix E1 in \cite{Betoule2014} to  bin the 1048 Pantheon SN Ia data points into 35 bins in  the $\log z$ space  (see Appendix \ref{Appendix:Binned_Pantheon} for details). 
The binned results and 79 low-redshift ($z<1.4$) GRBs calibrated by SN Ia are shown in Fig.~\ref{fig1}.  We must point out here that  although the redshift of Pantheon SN Ia spans to $z=2.26$, only the luminosity distances   at $z\leq1.4$ are used to calibrate the GRB. This is because there are only 6 SN Ia data points at $z>1.4$, which is too few to estimate the luminosity  distance  accurately.

With the improved Amati correlation as an example, the values of coefficients ($a$, $b$, $c$) in Eq.~(\ref{3_Amati}) can be obtained from the 79 low-redshift GRBs by maximizing the D'Agostinis likelihood~\citep{D'Agostini2005}:
\begin{eqnarray}\label{Lc}
		\mathcal{L}(\sigma_\mathrm{int},a,b,c)\propto\prod_{i} \frac{1}{\sqrt{\sigma_\mathrm{int}^2+\sigma_{y,i}^2+b^2\sigma_{x,i}^2}}
		\times\exp\left[-\frac{[y_i-y_\mathrm{copula}(x_i,z_i; a, b, c)]^2}{2\left(\sigma_\mathrm{int}^2+\sigma_{y,i}^2+b^2\sigma_{x,i}^2\right)}\right],
	\end{eqnarray}
where $\sigma_x$ and $\sigma_y$ are  the uncertainties of $x$ and $y$, respectively, and  $\sigma_\mathrm{int}$ is the intrinsic uncertainty of GRBs.
From  the error propagation equation, we find that $\sigma _y$ and $\sigma_x$  can be derived from Eqs~(\ref{Ep}) and (\ref{Eiso}) as  follows
\begin{eqnarray}\label{errepeiso}
	\sigma_y=\frac{1}{\ln10}\frac{\sigma_{E_\mathrm{iso}}}{E_\mathrm{iso}},~
	\sigma_x=\frac{1}{\ln10}\frac{\sigma_{E_{p}}}{E_{p}}
\end{eqnarray}
with
\begin{eqnarray}
	\sigma_{E_\mathrm{iso}}=4\pi d_L^2\sigma_{S_\mathrm{bolo}}(1+z)^{-1}.
\end{eqnarray}
Here $\sigma_{E_{p}}$ and $\sigma_{S_\mathrm{bolo}}$ are available in the observations of GRBs. In our analysis, the \textit{CosmoMC} code is used\footnote{The {\it CosmoMC} code is available at \href{https://cosmologist.info/cosmomc/}{https://cosmologist.info/cosmomc}.}. The obtained values of $\sigma_\mathrm{int}$ and coefficients  in Eqs.~(\ref{Amati}) and (\ref{3_Amati}) are shown in Tab.~\ref{Tab1}. From it, one can see that the differences of $\sigma_\mathrm{int}$ and $a$ between the standard and improved Amati correlations are negligible, while the difference of $b$ is very significant. Furthermore, we find that the redshift-dependent  correlation is favored since the value of $c$ deviates from zero at the $1\sigma$ confidence level (CL). 

\begin{deluxetable}{c|ccccc}
	\tablecaption{Constraints on the standard and improved Amati correlations from low-redshift calibration\label{Tab1}}
	\tablewidth{0pt}
	\tablehead{
		Correlations & $\sigma_\mathrm{int}$ & $a$ & $b$ & $c$ & $-2\ln\mathcal{L}$
	}
	\startdata
	Standard Amati & 0.511(0.047) & 52.722(0.063) & 1.295(0.135) & - & 121.819 \\
	Improved Amati & 0.509(0.045) & 52.869(0.144) & 1.194(0.149) & -0.232(0.206) & 120.475 \\
	\enddata
	\tablecomments{
		The unmarginalized best-fitted values with standard deviations of standard and improved Amati correlations from 79 low-redshift GRBs.
	}
\end{deluxetable}

\begin{figure}
	\gridline{
		\fig{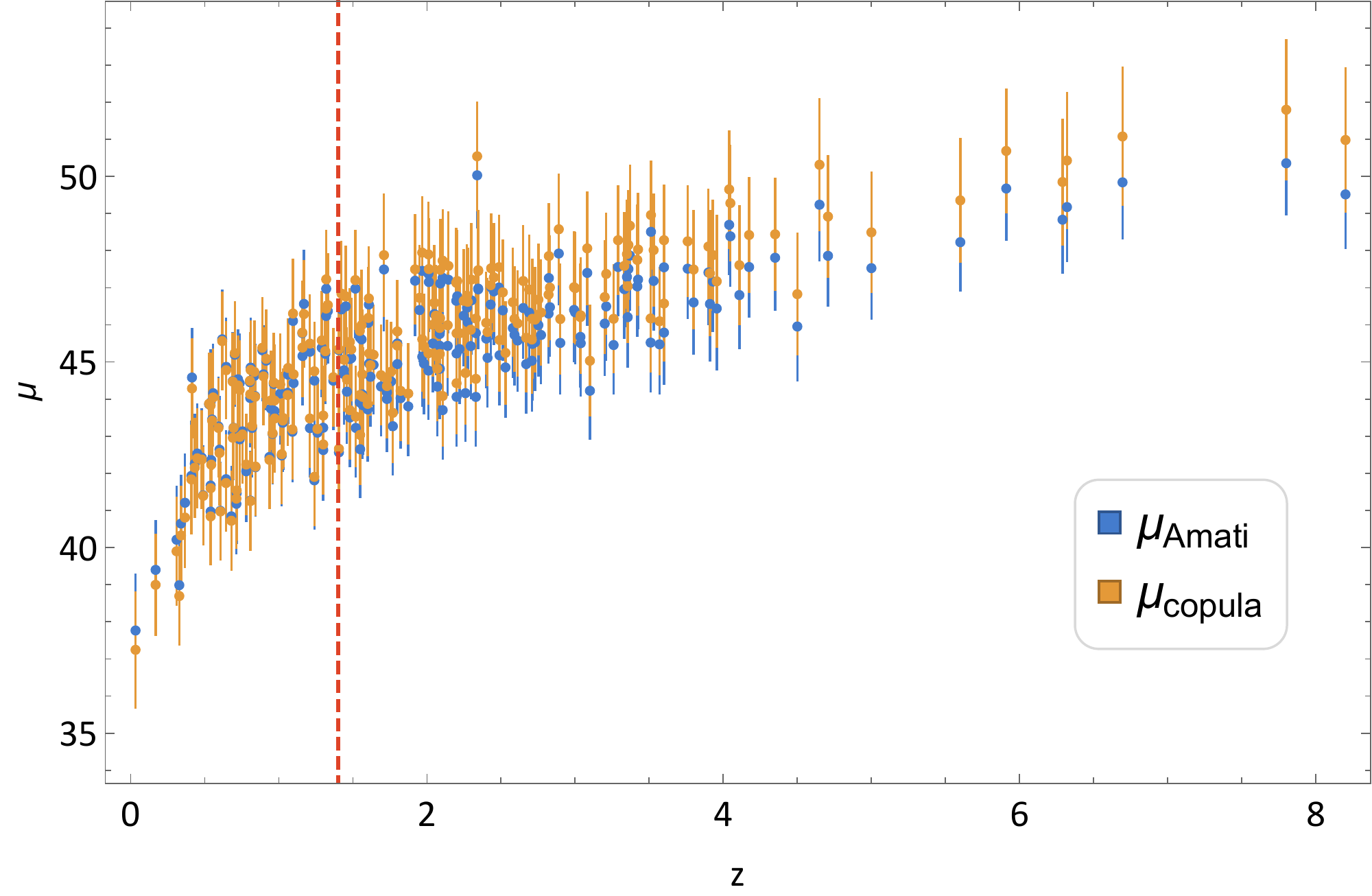}{0.5\textwidth}{(a)}
		\fig{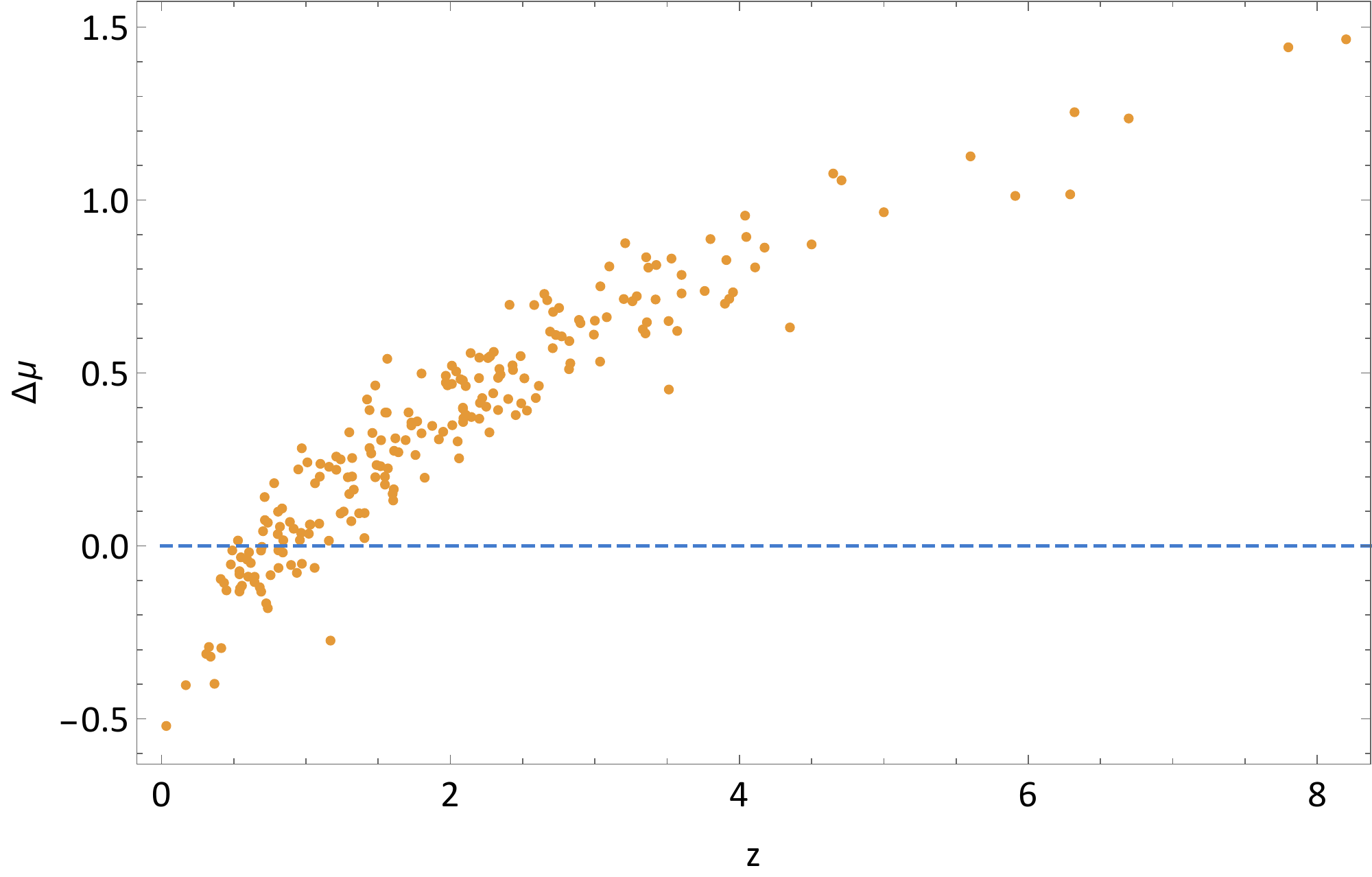}{0.5\textwidth}{(b)}
	}
	\caption{
		 The left panel is the Hubble diagram of 220 long GRBs calibrated from the standard and improved Amati correlations ($y_\mathrm{Amati}$ and $y_\mathrm{copula}$), respectively. The red dashed line denotes $z = 1.4$. The right panel shows the differences between $\mu_\mathrm{copula}$ and $\mu_\mathrm{Amati}$ ($\Delta \mu=\mu_\mathrm{copula}-\mu_\mathrm{Amati}$).}
	\label{fig:Hubble_diagram}
	
\end{figure}

Extrapolating the results from the low-redshift GRB data to the high-redshift one,   one can  derive the GRB luminosity distance from observations and obtain the corresponding distance modulus, which relates to the $d_L$  through 
\begin{eqnarray}\label{distance_modulus}
	\mu(z)=m-M=25+5\log\frac{d_L(z)}{\mathrm{Mpc}},
\end{eqnarray}
where $m$ and $M$ denote the apparent  and absolute magnitudes, respectively.
The uncertainty of the distance modulus is  propagated from the uncertainties of $S_\mathrm{bolo}$, $E_\mathrm{iso}$ and $E_p$:
\begin{eqnarray}\label{mu_err}
	\sigma_\mu^2=\left(\frac{5}{2}\sigma_{\log \frac{E_\mathrm{iso}}{1\mathrm{erg}}}\right)^2+\left(\frac{5}{2\ln10}\frac{\sigma_{S_\mathrm{bolo}}}{S_\mathrm{bolo}}\right)^2,
\end{eqnarray}
where
\begin{eqnarray}\label{errEiso}
	\sigma_{\log \frac{E_\mathrm{iso}}{1\mathrm{erg}}}^2&=&
	\sigma_\mathrm{int}^2+
	\left(\frac{b}{\ln 10}\frac{\sigma_{E_{p}}}{E_p}\right)^2+\sum_{i=1}^4\left(\frac{\partial y_\mathrm{copula}(x;\bm{\theta}_{c})}{\partial \theta_i}\right)^2 C_{ii}\nonumber\\
	&+&2\sum_{i=1}^4\sum_{j=i+1}^4\left(\frac{\partial y_\mathrm{copula}(x;\bm{\theta}_{c})}{\partial \theta_i}\frac{\partial y_\mathrm{copula}(x;\bm{\theta}_{c})}{\partial \theta_j}\right) C_{ij}.
\end{eqnarray}
Here $\bm{\theta}_c=\left\{\sigma_\mathrm{int},a,b,c\right\}$, and
$C_{ij}$ is the covariance matrix of these fitted coefficients.  We  construct the Hubble diagram of the GRBs and show it in Fig.~(\ref{fig:Hubble_diagram}a). The differences between $\mu_\mathrm{copula}$ and $\mu_\mathrm{Amati}$ are shown in Fig.~(\ref{fig:Hubble_diagram}b), which  indicates  that in the high-redshift region, the GRB distance modulus from the improved Amati correlation is larger apparently than that from the standard one.

\section{Constraints on cosmological models}\label{Sec:Cosmo&data}

The distance modulus of the A220 GRB data set can be used to constrain  cosmological 
models by minimizing the $\chi^2$:
\begin{eqnarray}\label{chi^2}
	\chi^2_\mathrm{GRB}=\sum_{i=1}^{N}\left[\frac{\mu_\mathrm{GRB}(z_i)-\mu_\mathrm{th}(z_i;\bm{p},\mu_0)}{\sigma_{\mu_i} }\right]^2,
\end{eqnarray}
where $N=220$ is the number of the GRB data, $\mu_\mathrm{GRB}(z_i)$ is the distance modulus of the GRB data at redshift $z_i$ and $\mu_\mathrm{th}$ denotes the theoretical value of the distance modulus, which is given by a cosmological model with $\bm{p}$ representing the model parameters. 
In Eq.~(\ref{chi^2}), $\mu_0=25-5\log H_0$, which  is a nuisance parameter and is  marginalized here by using the analytical method given in \citep{Nesseris2004}. Then,  the $\chi^2_\mathrm{GRB}$ is modified  to be
\begin{eqnarray}
	\tilde{\chi}^2_{\mathrm{GRB}}=A-\frac{B^2}{C} ,
\end{eqnarray} 
where
\begin{eqnarray}
	A&=&\sum_{i=1}^{N}\left[\frac{\mu_\mathrm{GRB}(z_i)-\mu_\mathrm{th}(z_i;\bm{p},\mu_0=0)}{\sigma_{\mu_i} }\right]^2,\nonumber\\
	B&=&\sum_{i=1}^{N}\frac{\mu_\mathrm{GRB}(z_i)-\mu_\mathrm{th}(z_i;\bm{p},\mu_0=0)}{\left(\sigma_{\mu_i} \right)^2},\nonumber\\
	C&=&\sum_{i=1}^{N}\left(\frac{1}{\sigma_{\mu_i} }\right)^2.
\end{eqnarray}

Except  the GRB data, we also consider 31 Hubble parameter measurements \citep{Stern2010,Moresco2012,Moresco2016,Zhang2014,Moresco2015,Ratsimbazafy2017,Ryan2018} spanning  the redshift from 0.07 to 1.965, which are determined by the cosmic chronometric technique~\citep{Loeb1998,Jimenez2002}.
For the $H(z)$ data set, the minimization of $\chi^2$ method is also applicable:
\begin{eqnarray}
	\chi^2_{H(z)}=\sum_{i=1}^{N}\left[\frac{H_\mathrm{obs}(z_i)-H_\mathrm{th}(z_i;\bm{p},H_0)}{\sigma_{H_i}^\mathrm{obs}}\right]^2.
\end{eqnarray}
Here $H_\mathrm{obs}(z_i)$ is the Hubble parameter measurement at redshift $z_i$ and $H_\mathrm{th}$ is the theoretical value of the Hubble parameter.
The constraints on cosmological models from  the $\mbox{GRB}+H(z)$ data can be  obtained by minimizing 
\begin{eqnarray}
	\chi^2_\mathrm{total}=\tilde{\chi}^2_\mathrm{GRB}+\chi^2_{H(z)}.
\end{eqnarray}

We consider two different cosmological models:  $\Lambda$CDM and $w$CDM.  The Hubble parameter $H(z)$ of  the $w$CDM as a function of redshift $z$ has the form
\begin{eqnarray}\label{Hubble_Parameter}
	\frac{H(z;\bm{p})^2}{H_0^2}=E^2(z;\bm{p})=\Omega_\mathrm{m0}(1+z)^3+(1-\Omega_\mathrm{m0})(1+z)^{3(1+w)},
\end{eqnarray}
which reduces to that of the $\Lambda$CDM when $w=-1$.  
Thus, we have   $\bm{p}\equiv\{\Omega_\mathrm{m0}\}$ for the $\Lambda$CDM and  $\bm{p}\equiv\{w,\Omega_\mathrm{m0}\}$ for the $w$CDM.  From Eq.~(\ref{Hubble_Parameter}), one can achieve the luminosity distance $d_L(z,\bm{p})$ 
\begin{eqnarray}\label{dl}
	d_L(z;\bm{p})=\frac{ (1+z)}{H_0} \int_{0}^{z} \frac{\mathrm{d}\tilde{z}}{E(\tilde{z};\bm{p})}.
\end{eqnarray}

\begin{figure}
	\includegraphics[width=0.45\textwidth]{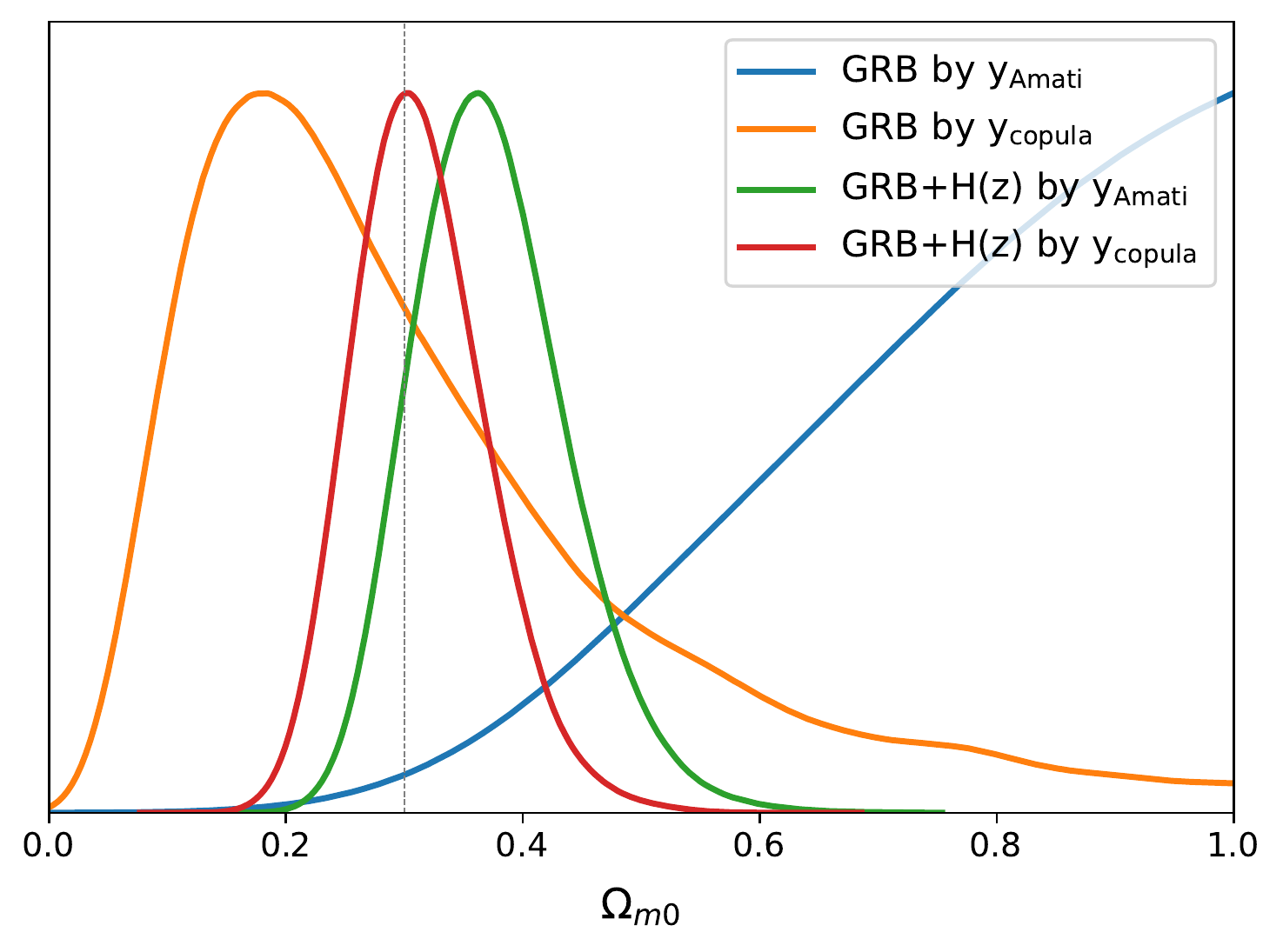}
	\includegraphics[width=0.45\textwidth]{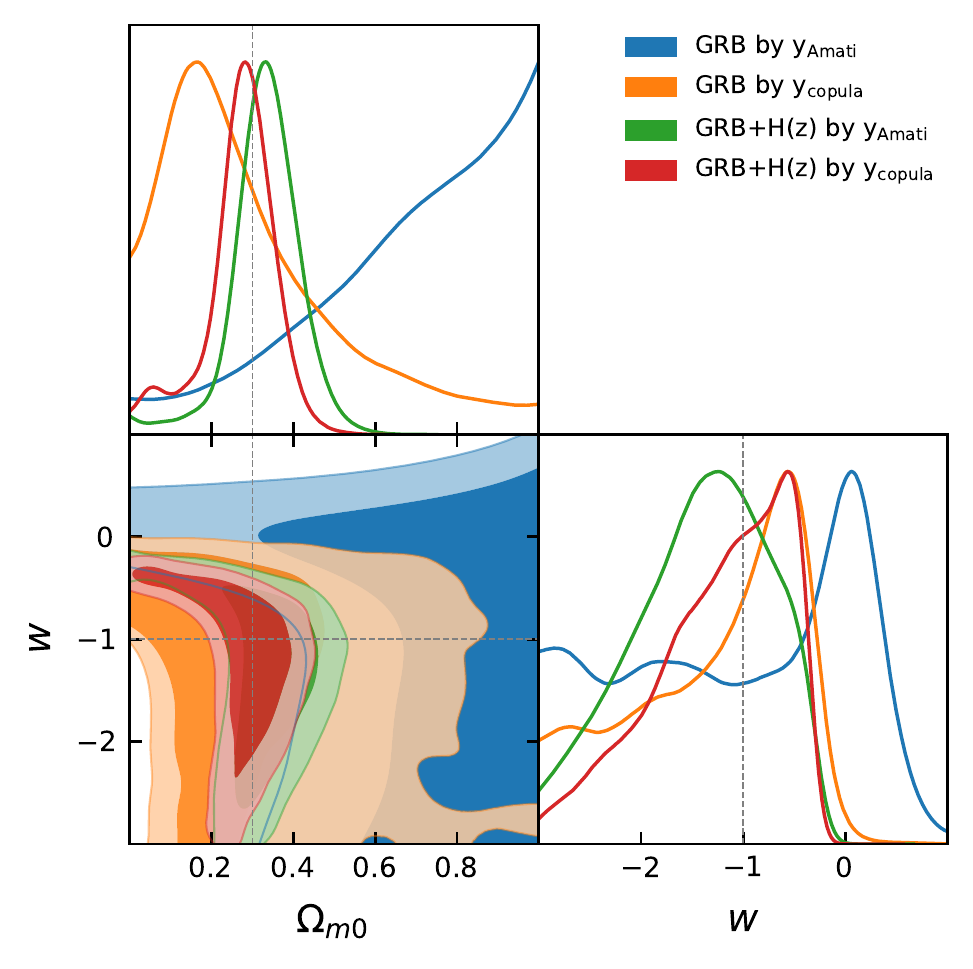}
	\caption{Constraints on the $\Lambda$CDM and the $w$CDM model from the GRB data and the $\mbox{GRB}+H(z)$ data.
		\label{fig:LCDM&wCDM}
	}
\end{figure}

The probability density distributions of the model parameters of two cosmological models are shown in Fig.~ (\ref{fig:LCDM&wCDM}), and the marginalized mean values with 68\% CL of these parameters are summarized in Tab.~\ref{tab2}. For a comparison, we consider both the standard and improved Amati correlations in standardizing the GRB. For the standard Amati correlation, it is easy to see that the GRB data only give a lower bound  limit on the  $\Omega_{\mathrm{m0}}$, which is similar to the results obtained in \citep{Khadka2021} with the simultaneous fitting method.  While, the GRB data from the improved Amati correlation can constrain  $\Omega_{\mathrm{m0}}$ more effectively, and  the mean values of $\Omega_{\mathrm{m0}}$  are $0.308^{+0.066}_{-0.230}$ and $0.307^{+0.057}_{-0.290}$ in the $\Lambda$CDM and $w$CDM models, respectively. These results are consistent very well with those given by other current popular observation data including BAO, CMB and so on~\citep{Aubourg2015,Scolnic2018,Planck}.
Furthermore, we find that for the $w$CDM model the GRB data from the standard Amati correlation only give an upper bound  limit  on $w$ at the $68\%$ CL, but those from the improved Amati correlation can give a tighter constraint.

\begin{figure}
	\includegraphics[width=0.45\textwidth]{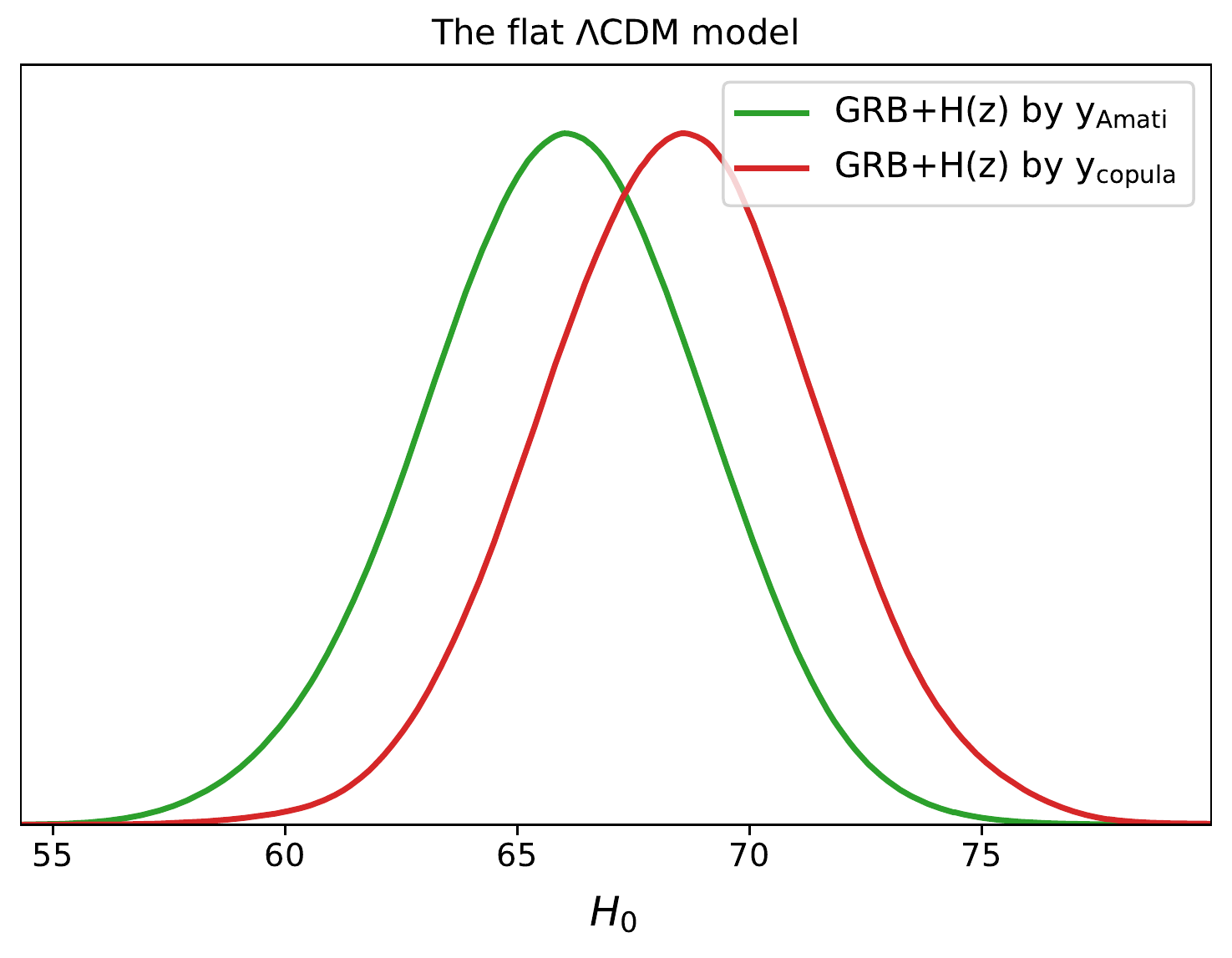}
	\includegraphics[width=0.45\textwidth]{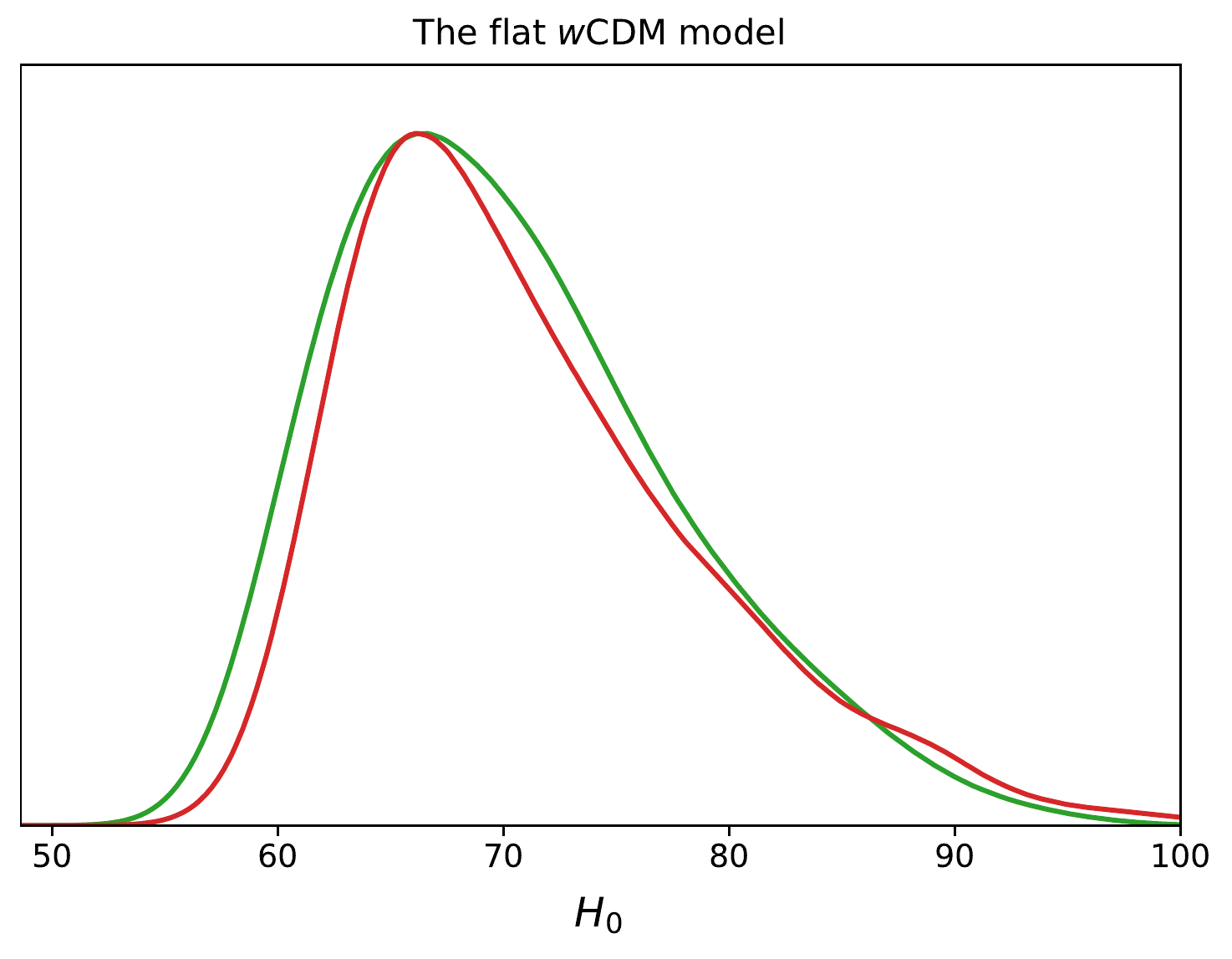}
	\caption{Constraints on $H_0$  from the $\mbox{GRB}+H(z)$ data. The green and red lines represent the results from the standard and improved Amati correlations, respectively. 
		\label{fig:marginal_H0}
	}
\end{figure}

When the $H(z)$ data are added in our analysis,  the $\mbox{GRB}+H(z)$ give tighter constraints    $\Omega_{\mathrm{m0}}$ and $w$ than those from the GRB  only.
We find that the GRB data from improved Amati correlation always favors a  smaller  $\Omega_{\mathrm{m0}}$ than that from the standard Amati correlation. The  $H(z)+$GRB from the improved Amati correlation can give a slightly tighter constraint on $w$ than the  $H(z)+$GRB from the standard  Amati correlation. We also investigate the limit on $H_0$ from  the $H(z)+$GRB. The results are shown in Fig.~(\ref{fig:marginal_H0}) and summarized in Tab.~\ref{tab2}. 
It is easy to see that two different correlations provide slightly different constraints on the Hubble constant, but the marginalized mean values seem to be close to the one ($67.4\pm 0.5~\mathrm{ km~s^{-1} Mpc^{-1}}$) from the Planck 2018 CMB observations~\citep{Planck}.

\section{Conclusions}\label{Sec:Conclusion}
An improved Amati correlation was constructed from the Gaussian \textit{copula} function recently  by us~\citep{Liu2022}. In this paper, we compare the constraints on cosmological models from GRB with the standard and improved Amati correlations. To obtain model-independently the GRB Hubble diagram, we  use the Pantheon SN Ia data  to calibrate the GRB data, and find that a redshift evolutionary correlation is favored slightly since the constant $c$, which is the coefficient of the redshift dependent term in the improved Amati correlation, deviates from the zero at the $1\sigma$ CL. The distance modulus of the GRB from the improved Amati correlation is apparently larger than  that from the standard Amati correlation at the high-redshift region ($z>1.4$). Thus, when the effect of the redshift evolution is neglected, the distance modulus of the GRBs will be underestimated at the high redshift region.

Using the GRBs to constrain the $\Lambda$CDM and $w$CDM models, we find that the data based on the standard Amati correlation only give a lower bound  limit  on $\Omega_\mathrm{m0}$ and an upper  bound one on $w$, while those from the improved Amati correlation can constrain $\Omega_\mathrm{m0}$ and $w$ more  tightly.  The mean values of $\Omega_{\mathrm{m0}}$ with the $68\%$ CL are $0.308^{+0.066}_{-0.230}$ and $0.307^{+0.057}_{-0.290}$ in the $\Lambda$CDM and $w$CDM models, respectively, which are consistent very well with those given by other current  popular observational data including BAO, CMB and so on~\citep{Aubourg2015,Scolnic2018,Planck}. When the $H(z)$ data are added together to constrain $\Lambda$CDM and $w$CDM,  tighter limits on $\Omega_\mathrm{m0}$ and $w$ are obtained.
Furthermore, a constraint on $H_0$ is achieved. We find that two different correlations provide marginally different $H_0$ results but the marginalized mean values seem to be close to that from the Planck 2018 CMB observations~\citep{Planck}.

\begin{deluxetable}{c|c|cccc}
	\tablecaption{Constraints on the $\Lambda$CDM and $w$CDM  from the GRB data  and the $\mbox{GRB}+H(z)$ data. \label{tab2}}
	\tablewidth{0pt}
	\tablehead{
		Correlations & Models & Data Sets & $H_0$ & $\Omega_{\mathrm{m0}}$ & $w$ 
	}
	\startdata
	& $\Lambda$CDM & A220 & - & $>0.651$ & - \\
	$y_\mathrm{Amati}$ & & A220+$H(z)$ & $66.195(3.105)^{+3.453}_{-3.139}$ & $0.370(0.067)^{+0.050}_{-0.079}$ & - \\
	\cline{2-6}
	& $w$CDM & A220 & - & $>0.573$ & $<-0.214$ \\
	& & A220+$H(z)$ & $70.176(7.701)^{+5.161}_{-9.516}$ & $0.333(0.085)^{+0.076}_{-0.068}$ & $-1.400(0.655)^{+0.864}_{-0.496}$ \\
	\hline
	& $\Lambda$CDM & A220 & - & $0.308(0.192)^{+0.066}_{-0.230}$ & - \\
	$y_\mathrm{copula}$ & & A220+$H(z)$ & $68.543(2.954)^{+2.861}_{-2.933}$ & $0.314(0.056)^{+0.046}_{-0.063}$ & - \\
	\cline{2-6}
	& $w$CDM & A220 & - & $0.307(0.219)^{+0.057}_{-0.290}$ & $-1.244(0.781)^{+1.016}_{-0.463}$ \\
	& & A220+$H(z)$ & $70.857(7.644)^{+4.310}_{-9.623}$ & $0.270(0.088)^{+0.090}_{-0.052}$ & $-1.203(0.625)^{+0.836}_{-0.327}$ \\
	\enddata
	\tablecomments{The marginalized mean values, the standard deviations, and the $68\%$ CL.
	}
\end{deluxetable}

\acknowledgments
This work was supported in part by the NSFC under Grants No. 12075084, No. 11690034, No. 11805063, No. 11775077, and 12073069,   
 by the Science and Technology Innovation Plan of Hunan province under Grant No. 2017XK2019, 
 and by the Guizhou Provincial  Science and Technology Foundation (QKHJC-ZK[2021] Key 020).

\appendix

\section{Binned distance modulus of Pantheon SN Ia}\label{Appendix:Binned_Pantheon}
Following the steps given in \citep{Betoule2014}, we use a piecewise linear function   to approximate the apparent magnitudes $m$ in the Pantheon SN Ia data, which is defined on each segment $z_i<z<z_{i+1}$ as
\begin{eqnarray}\label{binned_SN}
	\bar{m}(z)=(1-\alpha)~m_i+\alpha~m_{i+1}
\end{eqnarray}
with $\alpha=\log(z/z_i)/\log(z_{i+1}/z_i)$, where $m_i$ is the apparent magnitude at $z_i$. 
For the 1048 Pantheon SN Ia data points, we segment 35 bins with 36 $\log$-spaced control points   in the redshift region $0.01<z<2.3$.
To determine the value of $m_i$ at each control point $z_i$, the minimizing $\chi^2$ method is used:
\begin{eqnarray}
	\chi^2=[\bm{\hat{m}}(z)-\bar{m}(z)]^\dagger C^{-1} [\bm{\hat{m}}(z)-\bar{m}(z)].
\end{eqnarray}
Here $C$ is the covariance matrix with $1048 \times 1048$ elements, and $\bm{\hat{m}}$ is a one-dimensional array with 1048 observations in the Pantheon SN Ia sample. The results of 36 $\log$-spaced control points are shown in Tab.~\ref{tab3}. 
Since only the apparent magnitudes are obtained, we will set the absolute magnitude to be $M=-19.36$ \citep{Gomez-Valent2022} in our analysis to obtain the luminosity distance of GRBs.
If a different value of the absolute magnitude is chosen, it will change the value of coefficient $a$, but dose not affect the constraints on cosmological models.
 
\begin{deluxetable}{ccc|ccc|ccc}
	\tablecaption{Binned apparent magnitudes of the Pantheon SN Ia data. \label{tab3}}
	\tablewidth{0pt}
	\tablehead{
		$z_i$ & $m_i$ & $\sigma_{m_i}$ & $z_i$ & $m_i$ & $\sigma_{m_i}$ & $z_i$ & $m_i$ & $\sigma_{m_i}$
	}
	\startdata
	0.010& 13.912& 0.143& 0.065& 17.966& 0.049& 0.416& 22.469& 0.028\\
	0.012& 14.132& 0.131& 0.075& 18.293& 0.051& 0.486& 22.836& 0.030\\
	0.014& 14.604& 0.088& 0.088& 18.696& 0.040& 0.568& 23.272& 0.030\\
	0.016& 14.751& 0.059& 0.103& 19.094& 0.032& 0.664& 23.632& 0.041\\
	0.019& 15.208& 0.076& 0.120& 19.386& 0.026& 0.775& 24.081& 0.036\\
	0.022& 15.483& 0.053& 0.140& 19.825& 0.025& 0.905& 24.503& 0.041\\
	0.025& 15.839& 0.041& 0.164& 20.079& 0.025& 1.058& 24.893& 0.073\\
	0.030& 16.205& 0.041& 0.192& 20.530& 0.022& 1.235& 25.432& 0.117\\
	0.035& 16.550& 0.036& 0.224& 20.881& 0.023& 1.443& 25.570& 0.150\\
	0.040& 16.863& 0.047& 0.261& 21.231& 0.020& 1.686& 26.246& 0.224\\
	0.047& 17.234& 0.050& 0.305& 21.664& 0.021& 1.969& 26.130& 0.295\\
	0.055& 17.536& 0.044& 0.356& 22.042& 0.022& 2.300& 26.971& 0.297\\
	\enddata
	\tablecomments{The best-fitted value of binned apparent magnitude with standard deviation at each control point $z_i$.
	}
\end{deluxetable}

\end{document}